\documentclass[twocolumn,prl,superscriptaddress,showpacs]{revtex4}
\usepackage[colorlinks,bookmarks=false,citecolor=blue,linkcolor=red,urlcolor=blue]{hyperref}
\usepackage{graphicx,epstopdf,color}
\usepackage{amsfonts}
\usepackage{amsmath,amssymb,mathrsfs}
\usepackage{bm}
\usepackage{pst-grad}

\def\bc{\begin{center}}
\def\ec{\end{center}}

\newcommand{\bea}{\begin{eqnarray}}
\newcommand{\eea}{\end{eqnarray}}
\newcommand{\dagga}{{\phantom{\dagger}}}
\newcommand{\pd}{{\phantom{\dagger}}}

\def\ie{\emph{i.e.},\ }
\def\eg{\emph{e.g.}\ }
\def\ea{\emph{et~al.}}

\newcommand{\be}{\begin{equation}}
\newcommand{\ee}{\end{equation}}

\begin{document}
\title{Detecting Quantum Critical Points using Bipartite Fluctuations}
\author{Stephan Rachel}
\affiliation{Department of Physics, Yale University, New Haven, CT 06520, USA}
\author{Nicolas Laflorencie}
\affiliation{Laboratoire de Physique Th\'eorique, Universit\'e de Toulouse, UPS, (IRSAMC), F-31062 Toulouse, France}
\author{H. Francis Song}
\affiliation{Department of Physics, Yale University, New Haven, CT 06520, USA}
\author{Karyn Le Hur}
\affiliation{Department of Physics, Yale University, New Haven, CT 06520, USA}

\begin{abstract}
 We show that the concept of {\it{bipartite fluctuations}} $\cal F$ provides a very efficient tool to detect quantum phase transitions in strongly correlated systems. Using state of the art numerical techniques complemented with analytical arguments, we investigate paradigmatic examples for both quantum spins and bosons. As compared to the von Neumann entanglement entropy, 
we observe that $\cal F$ allows to find quantum critical points with a much better accuracy in one dimension. We further demonstrate that $\cal F$ can be successfully applied to the detection of quantum criticality in higher dimensions with no prior knowledge of the universality class of the transition. Promising approaches to experimentally access fluctuations are discussed for quantum antiferromagnets and cold gases.

\end{abstract}


\maketitle
Quantum phase transitions~\cite{vojta} occur at zero temperature and are solely driven by quantum fluctuations. Hence it is expected that a quantum phase transition should be manifested through the system's entanglement properties~\cite{osterloh-02n608-amico-08rmp517}.
Identifying appropriate measures of entanglement is, however, a non-trivial task. An important tool to access and quantify the amount of entanglement between two sub-sets $\cal A$ and $\cal B$ of an interacting quantum system is the von Neumann entanglement entropy (EE). In one dimension (1D), conformal field theory and exact calculations have established the logarithmic scaling of the von Neumann entropy\,\cite{Holzhey94-vidal-03prl227902-Calabrese04} for critical systems. For gapped systems the EE saturates to a constant and thus obeys a strict area law (assuming a local Hamiltonian)\,\cite{hastings07jsmp08024-eisert-10rmp277}.
In fact, EE can help to locate the quantum critical point (QCP) in some cases\,\cite{dalmonte}; for more subtile situations (\eg like Kosterlitz-Thouless transitions) it was demonstrated recently that the EE failed to locate the QCP of the frustrated $J_1$--$J_2$ chain\,\cite{alet-10prb094452}.
In higher dimensions, it was established that the gapless Heisenberg antiferromagnet (AF) on a square lattice obeys a strict are law\,\cite{kallin-09prl117203-hastings-10prl157201,song-11prb224410}, as also expected for a gapped phase. In such a situation, it is therefore unlikely that von Neumann EE will be a useful and practical tool to detect QCPs. Conversely, the valence bond entropy has been shown to be a powerful quantity to locate QCPs in any dimension, based on different scaling regimes, but it is restricted to SU(2)-invariant spin systems~\cite{Alet07,alet-10prb094452}. 

The aim of this Letter is to promote a general and more practical quantity to precisely locate QCPs for a larger variety of strongly correlated systems in any dimension $d$. Using the concept of {\it{bipartite fluctuations}}~\cite{Gioev06,song-10prb012405,song-11}  $\cal F$ of particle number or magnetization in many-body quantum systems, we focus on systems where such U(1) charges $\cal O$ are {\it globally} conserved while they {\it locally} fluctuate within each subsystems. We define 
\begin{figure}
\centering
\includegraphics[clip,width=\columnwidth]{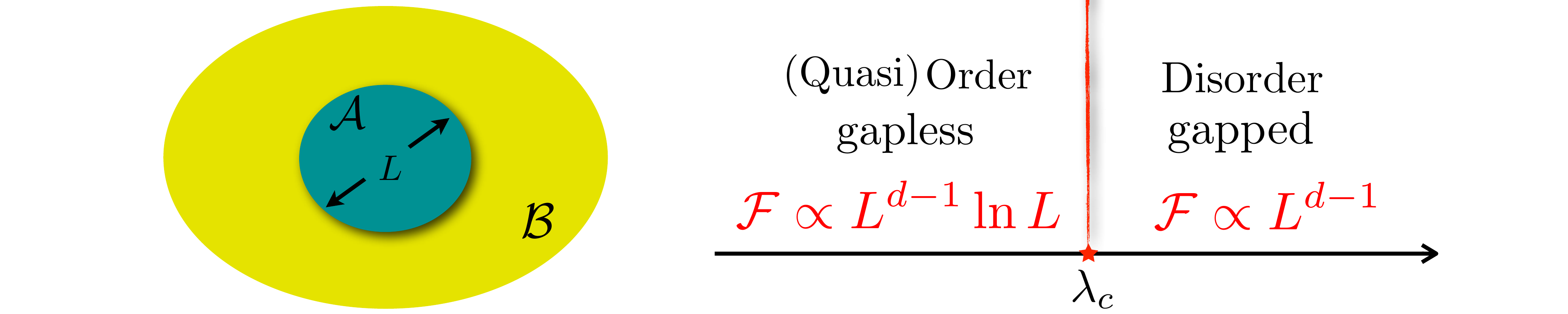}
\caption{(color online). For a $d$-dimensional system, the fluctuations ${\cal F}$ within subsystem $\cal A$ (of linear size $L$) with respect to $\cal B$ provide a precise estimate to locate a QCP at $\lambda_c$ between gapless (quasi) ordered and gapped disordered phases with disctinct scalings with $L$.}
\label{fig:schem}
\end{figure}
\be
{\cal{F_A}}=\Bigl\langle\Big(\sum_{i\in\cal A} {\cal O}_i\Big)^2\Bigr\rangle - \Bigl\langle\sum_{i\in\cal A} {\cal O}_i\Bigr\rangle^2,
\ee
where the (globally) conserved quantity ${\cal O}$ can be the particle number $n$ or the magnetization $S^z$ and $\langle\cdot\rangle$ refers to the ground state at $T=0$.
${\cal O}_i$ is defined for a sub-system $\cal{A}$ embedded in a larger one, see Fig.\,\ref{fig:schem}. For the special case that $\mathcal{A}$ is the total system, $\mathcal{F}_A$ is just the susceptibility (or compressibility, respectively) divided by temperature.
We show for various models, such as the spin-$\frac{1}{2}$ frustrated $J_1-J_2$ AF in 1D, the Bose-Hubbard chain at unit filling, 2D coupled Heisenberg ladders, and Bose-condensed hard-core bosons on a square lattice, that $\cal F_A$ provides a very efficient tool to accurately detect quantum criticality in the framework of quantum Monte Carlo (QMC) and Density Matrix Renormalization Group (DMRG) simulations on finite size systems\,\cite{suppl}. The key feature of the bipartite fluctuations is the distinct scaling behavior for gapless and gapfull phases in any dimension $d$ \cite{Gioev06,song-10prb012405}, as summarized in Fig.~\ref{fig:schem}:
within a sub-system of linear size $L$, $\cal F$ exhibits a strict area law for a disordered (gapped) ground-state,
$
{\cal F}_{\rm gapped}\propto L^{d-1}
$, 
whereas for a (quasi) ordered gapless state {\it{multiplicative}} logarithmic corrections appear, 
$ 
{\cal F}_{\rm gapless}\propto L^{d-1}\ln L
$, 
thus allowing to precisely locate a QCP between two such regimes. The bipartite fluctuations give an alternative view of the correlation functions since they are dominated by short-range fluctuations \cite{song-10prb012405}.
Experimentally, the concept of fluctuations has a very strong potential \cite{song-11}.

\vskip 0.15cm
{\it One dimensional systems---}
%
We now address 1D models, governed by Kosterlitz-Thouless (KT) type
quantum phase transitions usually difficult to precisely locate numerically. 
The first model we study
is the frustrated spin-$\frac{1}{2}$ $J_1$--$J_2$ chain, governed by the Hamiltonian 
\begin{equation}\label{ham:j1j2}
\mathcal{H}(\lambda)=\sum_i \left(\,{\bf{S}}_i\cdot{\bf{S}}_{i+1} +\lambda ~{\bf{S}}_i\cdot{\bf{S}}_{i+2}\,\right)\ ,
\end{equation}
where $J_2/J_1\equiv \lambda\ge 0$. For $\lambda\le \lambda_c$, this system has power-law critical correlations. At $\lambda_c\simeq 0.2412$, a KT transition into a dimerized phase occurs\,\cite{haldane82prb4925,okamoto-92pla433-eggert96prbR9612}. As mentioned above, the estimated value for the QCP using EE is not very precise compared to the established methods\,\cite{okamoto-92pla433-eggert96prbR9612} because the prefactor of the leading term in the EE (\ie the central charge $c$) is more or less insensitive to a change of $\lambda$ close to the QCP~\cite{alet-10prb094452}.
\begin{figure}
\centering
\includegraphics[clip,width=6.5cm]{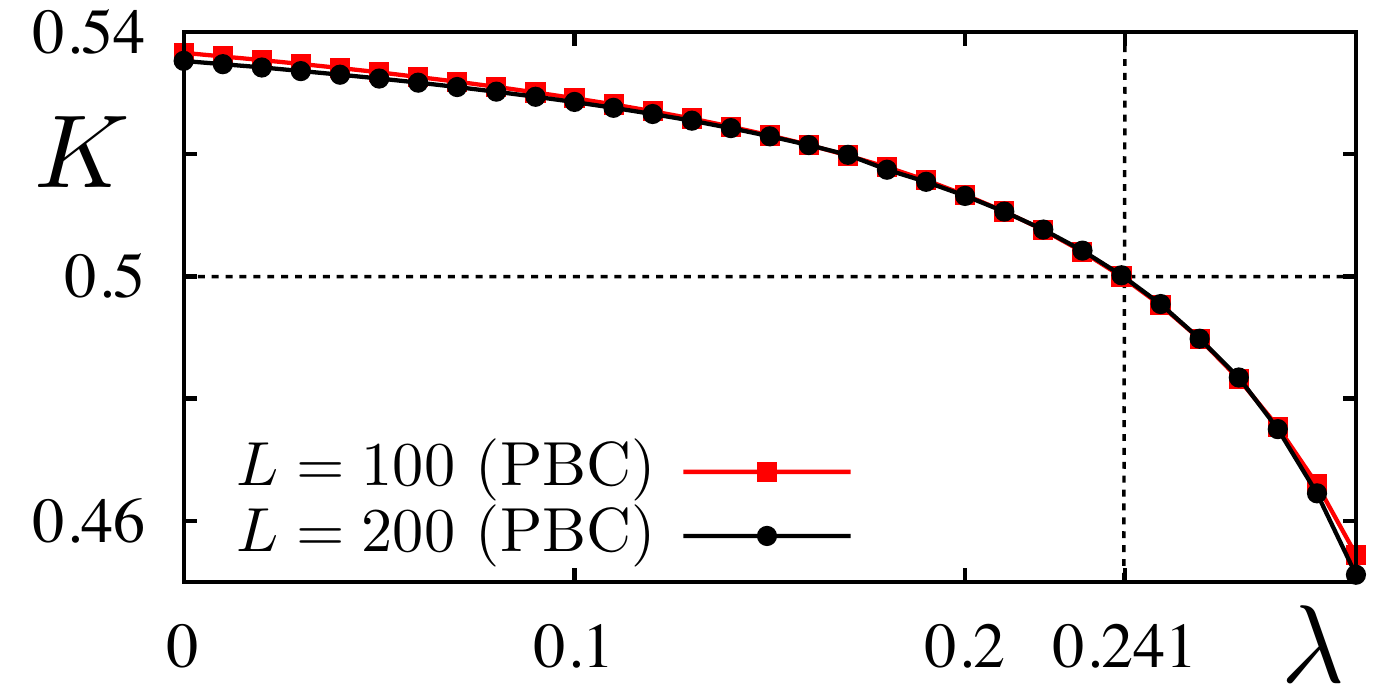}
\caption{(color online). Luttinger parameter $K$ of the $J_1-J_2$ chain extracted via \eqref{eq:F1D} vs.\ $\lambda\equiv J_2/J_1$. Shown is DMRG data for $L=100$ (red squares) and $200$ (black dots) for PBC.}
\label{fig:K-j1j2-pbc}
\end{figure}
Instead, we detect the transition by observing the behavior of $\cal F$ under variation of  the control parameter $\lambda$ which triggers the quantum phase transition. 
 The low-energy theory  describing such a quasi-ordered state is the Tomonaga-Luttinger liquid\,\cite{haldane82prb4925}, for which\,\cite{song-10prb012405,song-11}
\be
{\cal{F}}(L) = \frac{K}{\pi^2}\ln L +\rm cst,
\label{eq:F1D}
\ee
where $K=1/2$ is the Luttinger liquid parameter of this SU(2) point. However, marginally irrelevant operators lead to sizeable logarithmic corrections for $K$~\cite{Laflo01}, when computed on finite size systems. Interestingly, such corrections vanish precisely at $\lambda_c$ where $K$ quickly reaches its asymptotic value of $1/2$. Thus 
we have a systematic method at hand to detect this phase transition. In Fig.\,\ref{fig:K-j1j2-pbc}\, we have plotted the Luttinger parameter $K$ extracted from finite size DMRG calculations of Eq.~\eqref{eq:F1D}
versus $\lambda$. For PBC and $L=100$, $150$, $200$, and $250$, and after performing finite-size scaling, we obtain $\lambda_c= 0.2412(3)$ which agrees very well with the best estimates~\cite{okamoto-92pla433-eggert96prbR9612}. 
While there are a few other techniques available to find the QCP of the $J_1$--$J_2$ chain\,\cite{okamoto-92pla433-eggert96prbR9612,thomale-10prl116805,alet-10prb094452,roncaglia08,Thesberg11}, our approach stands out through efficiency and simplicity.

Another interesting model is the Bose-Hubbard chain: 
\begin{equation}\label{ham:bhm}
\mathcal{H}=-t\sum_{\langle ij \rangle} b_i^\dag b_{j}^{\phantom{\dag}} +
\frac{U}{2}\sum_i n_i \big(n_i - 1\big) -\sum_i \mu\, n_i\ ,
\end{equation}
where $t$ is the hopping amplitude, $U$ the on-site repulsion, and $\mu$ the chemical potential. 
Away from half filling, we expect a superfluid-Mott transition triggered by $\lambda\equiv t/U$. The superfluid phase is a Luttinger liquid\,\cite{Giamarchi04} with Luttinger parameter $K\ge 1$.
For unit filling, the QPT from a superfluid to a Mott insulator is of KT type (like in the $J_1-J_2$ chain discussed above). The complete $(\mu,t/U)$ phase diagram was carefully investigated within DMRG in Refs.\,\onlinecite{kuhner-98prb14741,kuhner-00prb12474,bhm1d}. Here we revisit the problem (restricted to unit filling) and show that we locate the transition with a better accuracy by virtue of the fluctuations. 
In the superfluid phase, the Green's function
$G(r)=\langle b_r^\dag b_0^{\phantom{\dag}}\rangle\propto r^{-1/2K}$
decays as a power-law. From Luttinger liquid theory we know that the transition occurs for $K_c=2$, see Ref.\,\onlinecite{giamarchi-92prb9325}. In Refs.\,\onlinecite{kuhner-98prb14741,kuhner-00prb12474} the Luttinger parameter $K$ was extracted directly from $G(r)$, thus giving an estimate of the critical point $\lambda_c=0.297\pm 0.01$~\cite{kuhner-00prb12474}. The major advantages of our approach is that (i) we have a finite size formula for the fluctuations (\ie applicability of conformal mappings) contrary to $G(r)$, and (ii) the computational cost of $\cal F$ using DMRG (see App.\,C of Ref.\,\onlinecite{song-11}) is much lower as compared to the Green's function at large distances. 
We extract $K$ from $\cal F$ for OBC with $L=64$, $128$, and $256$ (the latter is shown in Fig.3). By performing finite size scaling we obtain a much more precise estimate $\lambda_c = 0.2989(2)$, as compared to previous works~\cite{suppl}.
\begin{figure}
\centering
\includegraphics[clip,width=6.7cm]{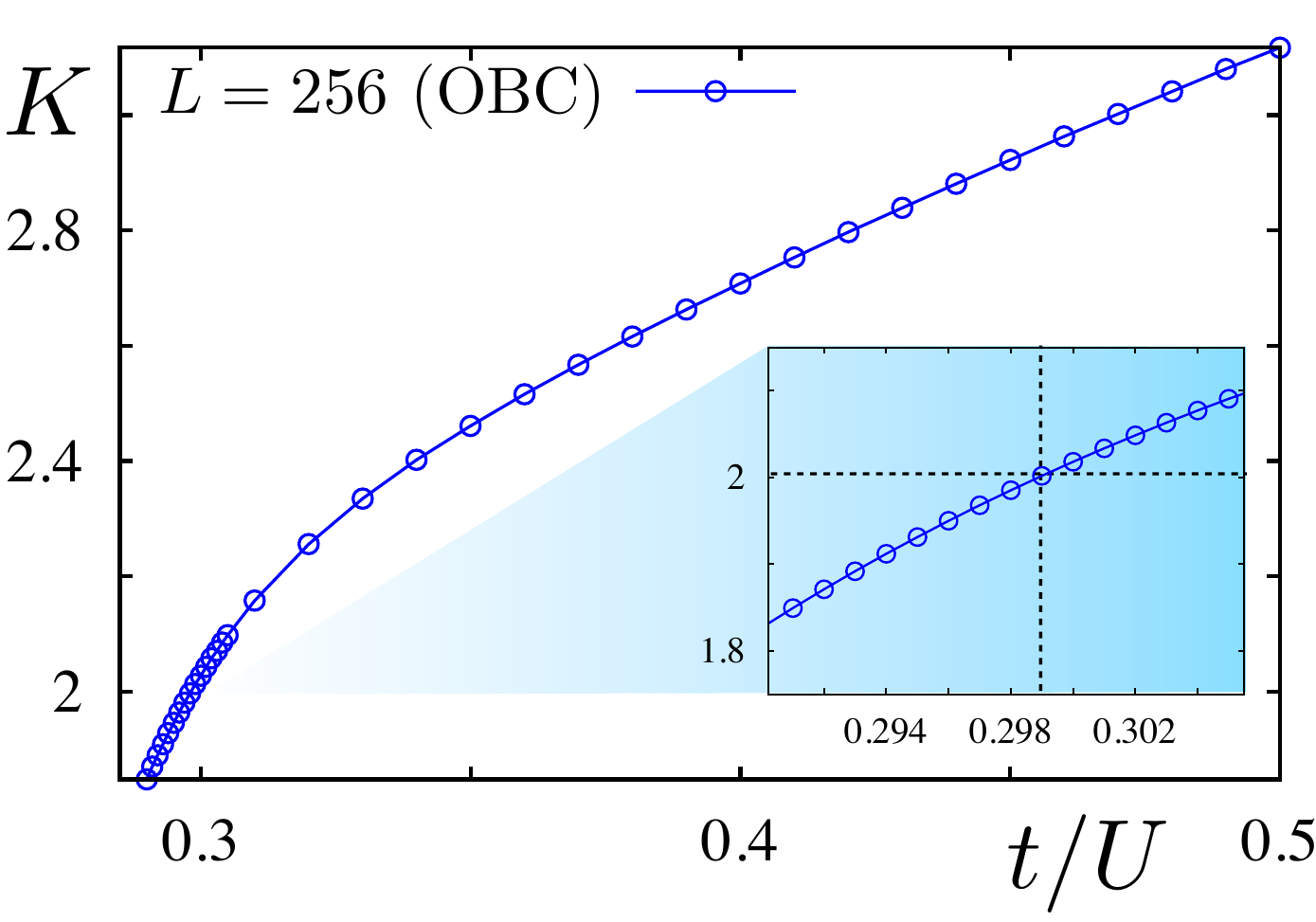}
\caption{(color online). Luttinger parameter $K$ of Bose-Hubbard chain extracted via \eqref{eq:F1D} vs.\ $\lambda\equiv t/U$ for $L=256$ (unit filling) and OBC. We restricted the local boson occupation number to 4\,\cite{kuhner-98prb14741,kuhner-00prb12474}. Inset: zoom close to the transition.}
\label{fig:K-bhm-obc}
\end{figure}
\vskip 0.2cm
{\it{Two dimensions---}}~Let us now move to 2D with a system of coupled spin-$\frac{1}{2}$ AF ladders, depicted in the inset of Fig.~\ref{fig:F2DHAF} (a), and governed by the Hamiltonian
%
\begin{figure*}[ht!]
\bc
\includegraphics[clip,width=15cm]{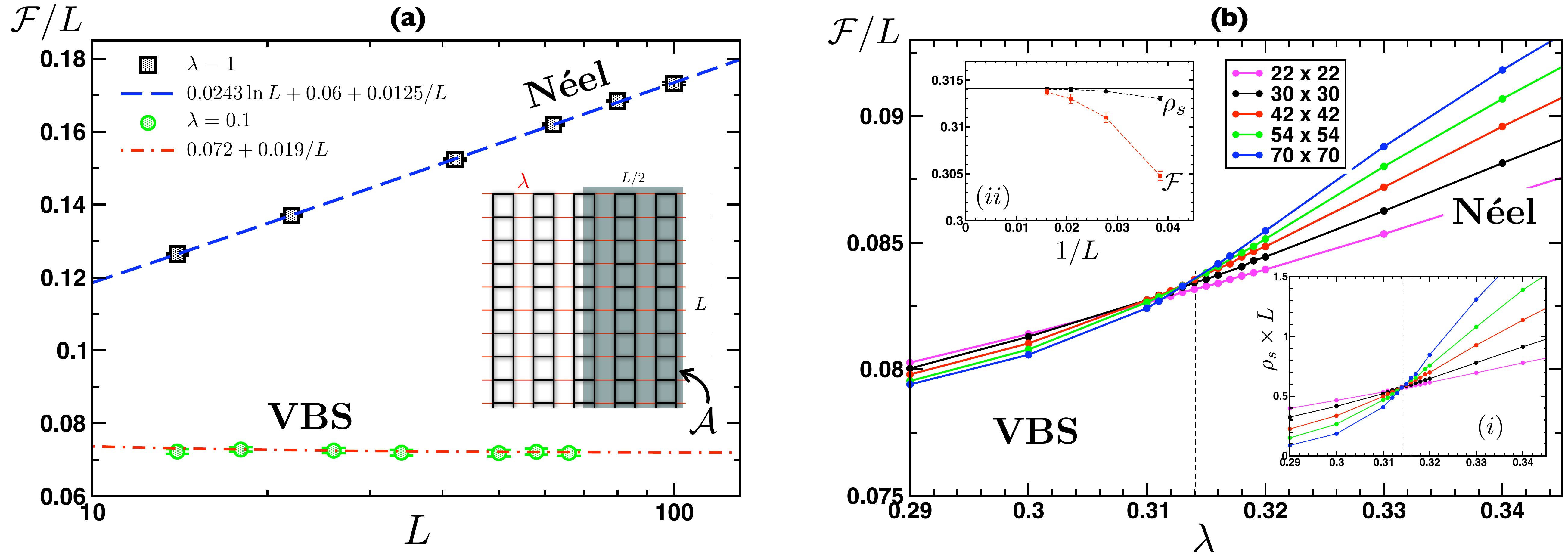}
\ec
\caption{(color online). Quantum Monte Carlo results for $T=0$ fluctuations $\cal F$ of the total magnetization in a region $\cal A$ for 2D coupled spin-$\frac{1}{2}$ ladders [Eq.~\eqref{eq:model}], depicted in the inset of (a). Left (a): ${\cal F}/L$ increases logarithmically with $L$ in the N\'eel regime (black squares $\lambda=1$) whereas it saturates to a constant in the valence bond state (green circles $\lambda=0.1$). Right (b): ${\cal F}/L$, plotted vs.\ $\lambda$ for various system sizes, displays a crossing point at $\lambda_c$. Insets: $(i)$ crossing of the stiffness $\rho_s\times L$ at $\lambda_c$ for the same sizes; $(ii)$ $1/L$ convergence of the crossing point for $\cal F$ (red squares) and $\rho_s$ (black circles) to the critical value (horizontal black line) $\lambda_c=0.31407$~\cite{Matsumoto02}.}
\label{fig:F2DHAF}
\end{figure*}
\be
{\cal H} = \sum_{\rm ladd.}\,\,{\bf{S}}_{i}\cdot{\bf{S}}_j \,\,\,+\sum_{\rm inter-ladd.} \lambda\,{\bf{S}}_{i}\cdot{\bf{S}}_j.
\label{eq:model}
\ee
This model~\cite{Matsumoto02,Wenzel08} displays a gapped valence bond solid (VBS) phase for small inter-ladder coupling $\lambda<\lambda_{c}$ with $\lambda_{c}=0.31407(5)$~\cite{Matsumoto02}, and a gapless N\'eel ordered phase for $\lambda>\lambda_c$.
Here we investigate the $T=0$ fluctuations of the total magnetization in a region $\cal A$ of size $x\times y$ embedded in a periodic square lattice $L\times L$. We choose a sub-system $\cal A$ with $x=L/2$ and $y=L$ which contains an even number of sites. 
QMC results for the $T=0$~\cite{comment} 
expectation of ${\cal F}(L/2)$ are shown in Fig. ~\ref{fig:F2DHAF}, with square lattices size up to $L\times L = 10^4$, for the isotropic square lattice $\lambda=1$ (N\'eel) and for weakly coupled ladders with $\lambda=0.1$ (VBS). In contrast with the entanglement (or R\'enyi) entropy which displays a strict area law in the N\'eel phase~\cite{kallin-09prl117203-hastings-10prl157201,song-11prb224410} (and presumably also in the VBS phase), the fluctuations follow a rather different scaling \cite{song-11prb224410}:
\be
\label{eq:F2D}
{\cal F}(\ell)\sim\left\{
\begin{array}{lr}
\alpha \ell\ln \ell +\beta \ell +\gamma &{\rm{~~{{(Gapless~NEEL)}}}}\\
\beta' \ell +\gamma'&{\rm{(Gapped~VBS).}}
\end{array}
\right.
\ee
Therefore, $\cal F/\ell$ plotted for different sizes will display a crossing point at $\lambda_c$, as we indeed observe in the panel (b) of Fig.~\ref{fig:F2DHAF} where the curves ${\cal F}(L/2)/L$ are plotted for various system sizes. 
The spin stiffness $\rho_s$, also known to be a useful quantity to locate a QCP, is shown in the right inset (ii) of Fig.~\ref{fig:F2DHAF} (b) where one sees a similar crossing for $\rho_s\times L^{d+z-2}$, with $z=1$ and $d=2$. As usual for such a technique, a drift of the crossing point is observed with $L$, as visible in the left inset (i) of Fig.~\ref{fig:F2DHAF} (b). Already known for a few other models~\cite{Wang06,Wenzel08}, the crossing points obtained from the stiffness converge very rapidly with $1/L$ to the bulk value $\lambda_c$, whereas we found a slower convergence for the estimates obtained from ${\cal F}/L$. Despite such effect (which may not be generic but model dependent), we demonstrate here with this simple example that $\cal F$ is a very useful quantity to locate a QCP between ordered and disordered phases for $d>1$. 

\begin{figure}[ht!]
\centering
\includegraphics[clip,width=\columnwidth]{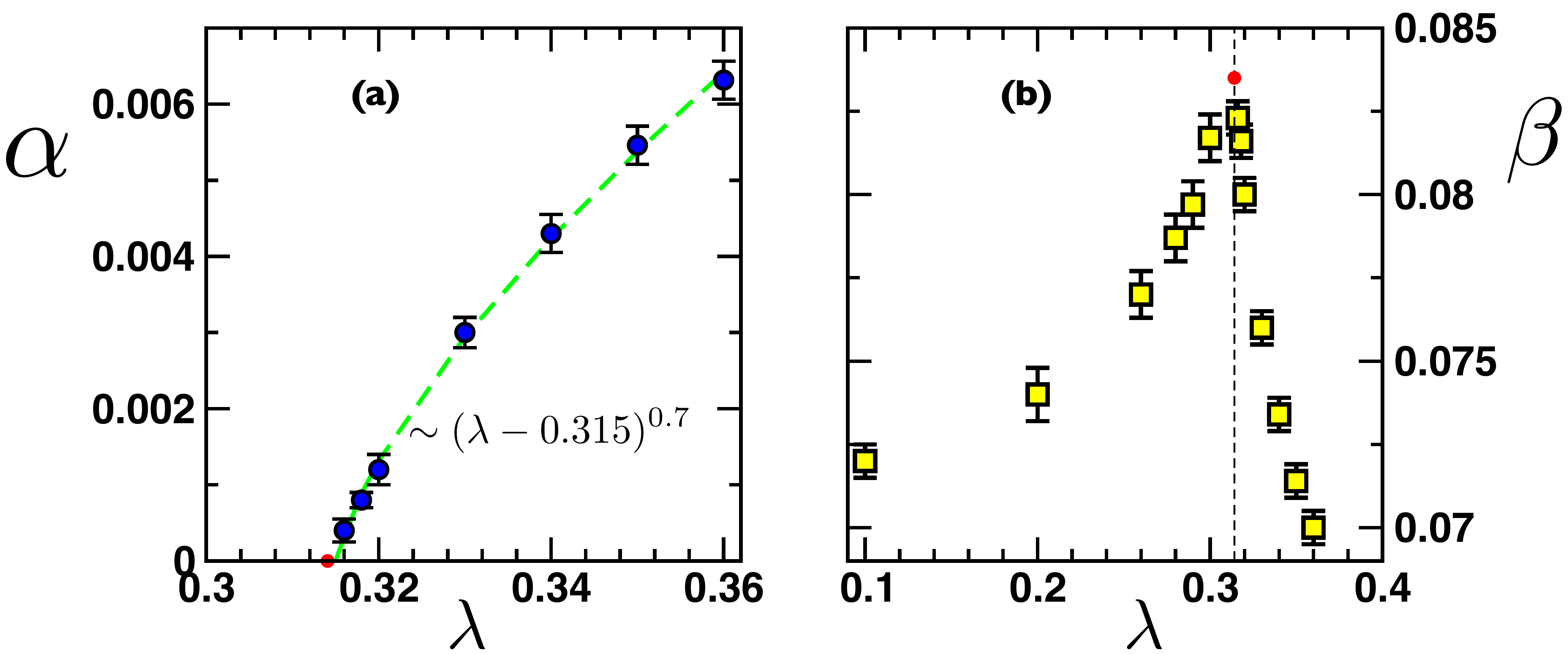}
\caption{Prefactors $\alpha$ and $\beta$ from Eq.~\eqref{eq:F2D} for coupled Heisenberg ladders, extracted from QMC data of Fig.~\ref{fig:F2DHAF}, and plotted against $\lambda$. (a) The critical point is shown by a red circle, and the green curve is the power-law fit indicated on the plot. (b) The vertical dashed line signals the critical coupling $\lambda_c$ and the crossing point (red circle) is at $\beta_c\simeq 0.0835$.}
\label{fig:alpha_and_beta}
\end{figure}
%

One can get even more insight from the behavior of the coefficients $\alpha$ and $\beta$ in Eq.~\eqref{eq:F2D} as a function of the inter-ladder coupling $\lambda$ (see Fig.~\ref{fig:alpha_and_beta}). The prefactor $\alpha$ of the leading term $\sim L\ln L$ in the N\'eel phase  vanishes at the critical point $\alpha\sim (\lambda-\lambda_c)^{x}$,
with $x\simeq 0.7$ and $\lambda_c= 0.315(1)$, in good agreement with the value $0.31407(5)$~\cite{Matsumoto02}. The area law term $\beta L$, displayed in  Fig.~\ref{fig:alpha_and_beta} (b), although certainly non-universal, exhibits a very interesting  $\lambda$-shape and passes through a maximum $\beta_c\simeq 0.0835$ at the critical coupling $\lambda_c$.

\begin{figure}
\centering
\includegraphics[clip,width=\columnwidth]{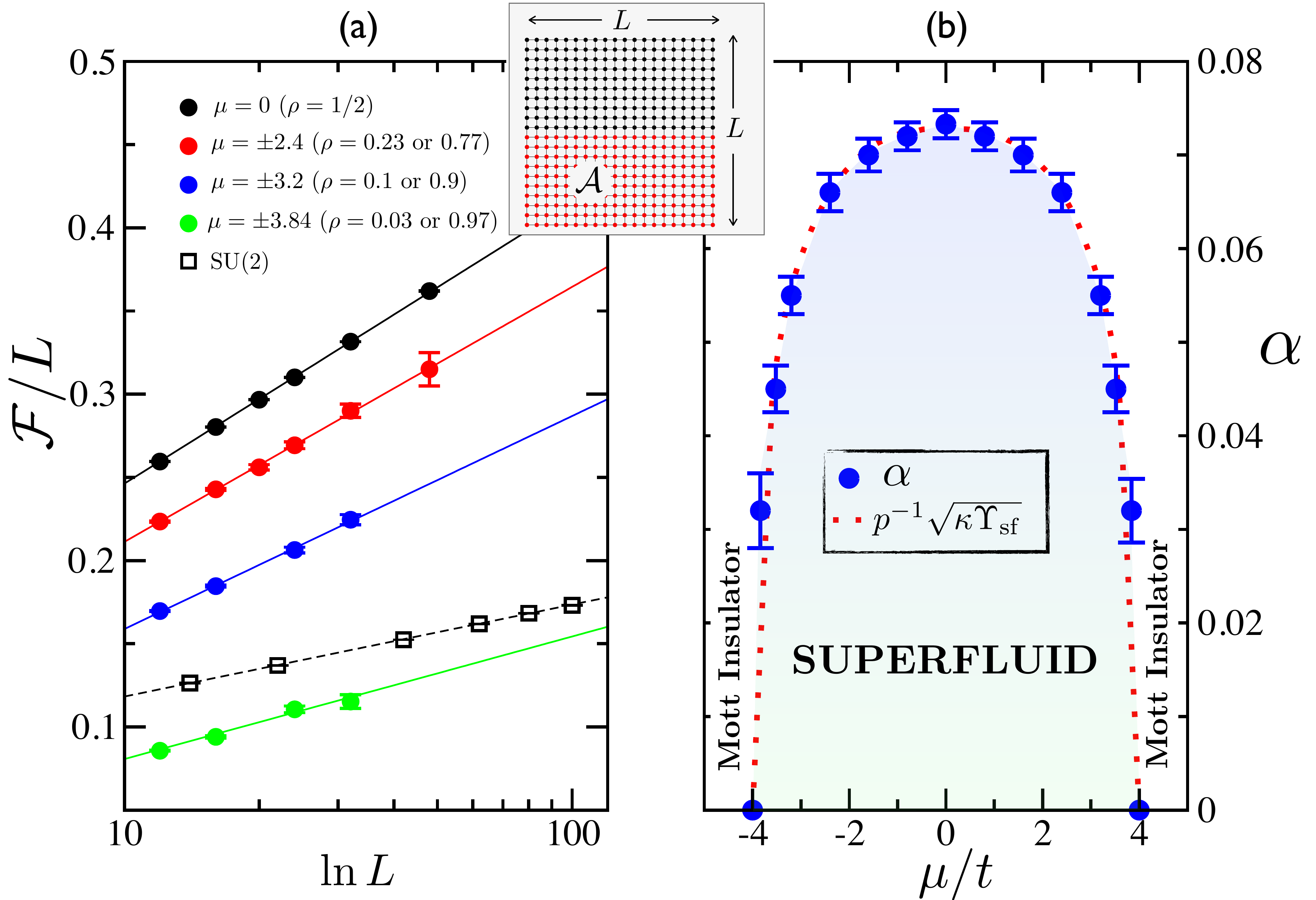}
\caption{(color online). QMC results for the fluctuations $\cal F$ of the particle number for hard-core bosons on square lattices of size $L\times L$, the subsystem $\cal A$ being half the lattice (see inset). (a) ${\cal F}/L$ plotted against $\ln L$ for 4 representative fillings indicated on the plot, and compared to the SU(2) Heisenberg model. (b) Prefactor $\alpha$, obtained from fits to the form Eq.~\eqref{eq:F2D}, shown over the entire regime versus $\mu/t$ together with $\sqrt{\kappa\Upsilon_{\rm sf}}/p$ from QMC, with $p=3.2(1)$.}
\label{fig:HCB}
\end{figure}

It is important to emphasize that, contrary to the stiffness, a prior knowledge of any critical exponent, such as the dynamical exponent $z$, is not necessary to precisely locate the QCP. Note also that we expect the valence bond entropy~\cite{Alet07} to display similar crossing properties for such a SU(2) symmetric Hamiltonian Eq.~\eqref{eq:model}. In order to illustrate further the general character of this method, we focus now on a non-SU(2) model: hard-core bosons on the square lattice. Governed by the Hamiltonian
\be
{\cal H}=-t\sum_{\langle i j\rangle}\left(b_{i}^{\dagger}b_{j}^{\dagga}+ {\rm{h.c.}}\right) -\mu\sum_i b_{i}^{\dagger}b_{i}^{\dagga},
\label{eq:hcb}
\ee
where $b$ are hard-core bosons operators, $t$ the hopping integral and $\mu$ the chemical potential, hard-core bosons on the square lattice~\cite{Bernardet02} exhibit a  particle-hole symmetric phase diagram at $T=0$ with a Bose-condensed superfluid state for $|\mu/t|<4$, and trivial Mott insulating phases for $|\mu/t|>4$, the transition between them being in the universality class of the diluted Bose gas with $z=2$. The Bose-condensed (U(1)-broken-symmetry) state is expected to display for $\cal F$ (fluctuations of the particle number) a similar scaling as the one observed for SU(2)-broken N\'eel-ordered spins, whereas for the trivial Mott insulators we simply have ${\cal F}_{\rm Mott}=0$. In Fig.~\ref{fig:HCB} (a) $T=0$ QMC results obtained for $\mathcal{F}$ are shown for 4 representative values of the chemical potential. The prefactor $\alpha$ of the $L\ln L$ term is plotted versus the chemical potential $\mu$ in the right panel of Fig.~\ref{fig:HCB} where we observe a very interesting dome-like shape in the superfluid regime. One can use an interesting analogy with quasi-one dimensional systems where the Josephson type interchain tunneling term  will lock the superfluid phase difference
between all chains. The low-energy (quasi-ordered) superfluid phase is described in terms of a single macroscopic 1D gapless mode.  For a number of chains $N=L$ then we predict ${\cal F}=(KL/\pi^2)\ln L$.
The logarithmic scaling of ${\cal F}$ is controlled by the Luttinger parameter $K$ of the effective theory. In the hydrodynamic description of a Luttinger liquid $K=\pi\sqrt{\kappa\Upsilon_{\rm sf}}$, where $\kappa$ is the compressibility and $\Upsilon_{\rm sf}$ is the stiffness. 
This gives $\alpha=\sqrt{\kappa\Upsilon_{\rm sf}}/\pi$. A similar  quantum-hydrodynamic theory for interacting bosons is obtained in two dimensions using the Gross-Pitaevskii approach. Comparing the prefactor $\alpha$ with $\sqrt{\kappa\Upsilon_{\rm sf}}$ (obtained in the same QMC simulation), as shown in Fig.~\ref{fig:HCB} (b), gives a very good agreement. We find the following result for the entire superfluid regime: $\alpha(\mu) =\sqrt{\kappa \Upsilon_{\rm sf}}/p$ with a coefficient $p\simeq 3.2(1)$. Scaling relations close to a QCP at $\lambda_c$ predict $\Upsilon_{\rm sf}\sim \xi^{2-d-z}$ and $\kappa\sim\xi^{z-d}$, thus leading to $\alpha\sim (\lambda-\lambda_c)^{\nu}$, which can be compared to Fig.~\ref{fig:alpha_and_beta} (a) where the exponent $x\simeq 0.7$ is very close to $\nu=0.709(6)$ of the 3D Heisenberg universality class~\cite{Wenzel08}.

\vskip 0.15cm
{\it{Conclusion---}} The concept of bipartite fluctuations of a (strongly correlated) quantum system has been shown for various paradigmatic condensed matter models to be an efficient, accurate, and rather general tool to detect quantum critical points
using state of the art numerical techniques.
In contrast to the von Neumann entropy, the fluctuations can be successfully used even in two spatial dimensions to find the critical point. Promising paths to directly measure the fluctuations have been proposed recently\,\cite{song-11}; particularly interesting proposals are quantum magnets in an external magnetic field with Meissner screens (covering region $\mathcal{B}$) as well as direct measurement of $\mathcal{F}$ using single atom microscopes \cite{exp-coldatom}. A next step will be to test the usefulness of this tool for unconventional quantum criticality~\cite{Sandvik07}.

\begin{acknowledgements}
SR acknowledges support from DFG under Grant No.\ RA~1949/1-1 and partially from NSF DMR 0803200, and thanks P.\ Schmitteckert for use of his DMRG code. HFS and KLH acknowledges support from NSF DMR 0803200 and from the Yale center for Quantum Information Physics
through NSFR DMR-0653377.
\end{acknowledgements}

\newpage
\phantom{x}
\newpage

\begin{widetext}
\section*{\Large Supplementary Material for\\
``Detecting Quantum Critical Points using Bipartite Fluctuations''}
\end{widetext}
%
%
%

Bipartite fluctuations\,\cite{long} bring a very strong concept in the study of quantum criticality.
Our approach to locate Quantum Critical Points stands out through the combination of the attributes \emph{generality}, \emph{simplicity}, \emph{efficiency}, and \emph{accuracy}. 
While other existing methods might be comparable with regard to one of these attributes, none of them is comparable with regard to all of these attributes as we shall show in the following.
While it would be an impossible task to prove this statement for any known quantum critical point within any numerical approach, we rather focus here on a representative example and leave it to the reader to convince himself/herself about the usefulness of our method.
We show below for the Kosterlitz-Thouless transition between superfluid and Mott insulator phases of the Bose-Hubbard chain at unit filling (numerical hard to study due to finite size logarithmic corrections)  that the comparison between our approach and several existing ones is unambiguously in favor of our method.

\section{Generality}
Our method is applicable to spins and interacting bosons (both demonstrated in the main text) but also to interacting fermions. It works equally well in any dimension (demonstrated for $d=1,2$ in the main text). Apart from the U(1) symmetry of spin or charge (which is of course essential for our approach), presence or absence of any symmetry such as \eg SU(2) does influence the precision of our method. Moreover, no prior knowledge about the order parameter or any critical exponent is required to locate the quantum critical point. Finally, our approach even enables us to locate quantum critical points of Kosterlitz--Thouless type (as demontrated for two different models in $d=1$), notoriously cumbersome due to finite size logarithmic corrections.

\section{Simplicity}
The fluctuations can be easily computed for a given subsystem, numerically and often even analytically. For numerical approaches where the reduced density matrix of a subsystem is computed the bipartite fluctuations are a side--product (see App.\,C of Ref.\,\onlinecite{long}). Examples are the density matrix renormalization group (DMRG) and its descendants. 
For other numerical methods, $\mathcal{F}_A$ requires to compute diagonal correlation functions $G(i,j)=\langle S_i^z S_j^z\rangle$ for magnets or $\langle n_i n_j\rangle$ for itinerant systems.
Simple crossing technique, \ie plotting $\mathcal{F}/L^{d-1}$ vs.\ $\lambda$  for different system sizes $L$ reveals the quantum critical point.

\section{Efficiency}
Our approach is very efficient for the DMRG technique where the reduced density matrix can be directly computed. For Quantum Monte Carlo techniques, while $\mathcal{F}_A$ is not conserved during the propagation along the Trotter direction, one can easily keep track of its evolution without computing all correlators at each imaginary time slice. As a result, the computational cost for estimating ${\cal F}$ is the same as of getting either the stiffness or a given structure factor. If the type of order is unknown, the full set of correlations will be required, which is, computationally speaking, much more expensive. In such a case, computing $\cal F$ would be far much easier.

Alternatively, one could compute the gap for various system lengths, but then one needs to perform a finite size scaling analysis which involves a prior knowledge of the universality class of the transition. Moreover, an efficient estimate of a tiny gap with QMC or DMRG (also discussed below) is computationally more costly than the other observables discussed above.

If one knows which kind of order the system will achieve, the natural computation would be the associated structure factor, with the complication that finite size scaling at the transition may be tricky to analyze. Therefore one usually prefers to rely on crossing techniques. For instance the order parameter squared (structure factor) $S\times L^{2\beta/\nu}$ will displays a crossing at a QCP, but one needs to know the ratio $\beta/\nu$ of two important critical exponents. The stiffness $\rho_s$ is usually better since it only requires to know one exponent: the dynamical exponent $z$. The good thing with the crossing of the stiffness is that subleading finite size corrections are usually very small (see for instance Ref.\,\onlinecite{wang06})
and therefore the location of $\lambda_c$ is very good. Nevertheless, in some case the dynamical exponent is an important unknown quantity of the problem and the crossing of $\rho_s\times L^{2-d-z}$ will be useless (here $d$ is the spatial dimension). As a consequence, the $\cal F$-crossing technique, which only relies on the result that ${\cal F}(x)\sim x^{d-1}\ln x$, provides a very nice and alternative way without any assumption regarding the critical point universality class.

In the following 
we consider the one-dimensional Bose--Hubbard model and use the DMRG method. First, we compute the fluctuations for a given parameter setting and, second, the Green's function $G(r)=\langle b^\dag_0 b^\pd_r\rangle$ for the same setting. We eventually compare the used CPU times.

\begin{figure}[t!]
\centering
\includegraphics[scale=0.75]{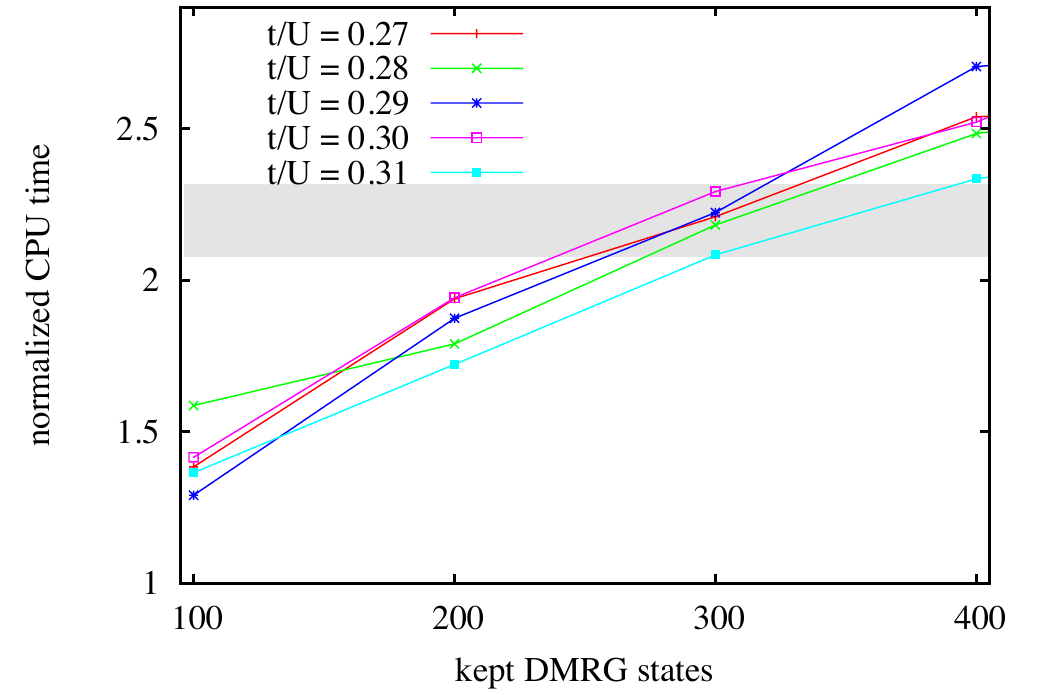}
\caption{Normalized CPU time using a single core vs.\ kept DMRG states for Bose--Hubbard chain, $L=128$, OBC. For about 300 kept DMRG states convergency is guaranteed implying a normalized CPU time $>2$.}
\label{fig:cpu}
\end{figure}

CPU time depends on many internal (\ie method-specific) parameters of the DMRG method. We tried to use a standard setting which is representative for most DMRG implementations. We always computed the fluctuations and the correlation functions on the same computer using a single core (in order to rule out the influence of the parallelized code). We also kept the same parameter setting for all computations. In addition, we varied the number of kept DMRG states (which can be converted into a measure for convergency) and for different interaction strengths $t/U$. Since computations for different values of $t/U$ have been performed on different computers, we normalized the CPU time, 
$$\hbox{normalized CPU time}  = \frac{\hbox{CPU time for correlations}}{\hbox{CPU time for fluctuations}}\ .$$
For 300 kept DMRG states we found the ``discarded entropy sum'' to be smaller than $10^{-10}$ which guarantees convergency within the DMRG method (we always performed 15 sweeps). For all considered values $t/U$ (see Fig.\,\ref{fig:cpu}) the normalized CPU time is between 2.1 and 2.3, \ie it takes more than the double time to compute correlations compared to fluctuations. In this example, we chose the system length to be $L=128$ sites and we used OBC. In the main text, we also computed $L=256$ where the normalized CPU time is even higher. Below we further demonstrate that $\cal F$ is the best quantity to easily locate the critical point of the 1D Bose- Hubbard model, as presented in the main text.

Before this, let us briefly comment on investing the gap size as an alternative method to locate the quantum critical point. For most numerical methods, the performance to compute the gap size between ground state energy and excitation spectrum is rather bad. In the DMRG method, for instance, the normalized CPU time would easily exceed a value of 4 to 5.

\section{Accuracy}

For the crossing technique, \ie plotting $\mathcal{F}/L^{d-1}$ vs.\ $\lambda$ for different system sizes $L$, one only needs to know the value of the fluctuations when subsystem $A$ is \eg half the total system length. No fitting process is involved (in contrast to correlation functions (see below)). The crossing technique works extremely well as demonstrated in our paper for both two-dimensional models. In one dimension, it works as well, but here in order to beat the existing approaches (such as level-spectroscopy of the full energy spectrum) we used a somewhat more sophisticated way and extracted the Luttinger parameter $K$ from the fluctuations. This involves a fitting process but due to the availability of conformal mappings we have a finite size formula at hand which simplifies the fitting process drastically and makes it much less dependent of the fitting window. In contrast, extracting $K$ from the correlation function appears to be a much more delicate procedure. 
For the Green's function $G(r)$ we {{still}} have a finite size formula\,\cite{cazalilla04},
$$\langle b^\dag_0 b^\pd_r\rangle\!=\! G_0 \frac{\pi}{2L}\!\!\left(\!\frac{\sqrt{\sin(\pi r/L)\sin(\pi/L)}}{\sin(\pi(r+1)/(2L))\sin(\pi(r-1)/(2L))}\!\right)^{\!\frac{1}{2K}}$$
but the extracted value of $K$ is very sensitive to the fitting window which is used. This is exemplified in the left panel of Fig.\,\ref{fig:corr-fluc} below, where DMRG results are shown for an open Bose-Hubbard chain of $L=128$ site, in the superfluid regime ($n=1$, $t/U=0.3$), but close to the Mott insulating phase.
Indeed, the extracted value of the Luttinger liquid parameter $K$ strongly depends on the fitting process, depending on the type and the size of the retained fitting window. Consequently the estimated $K=1.93(11)$ displays a quite large uncertainty, as already discussed in T.~D. K\"uhner, S.~R. White, and H. Monien (Ref.\,\onlinecite{Kuhner00}). In 
contrast, the bipartite fluctuation provide a much more accurate tool, as shown in the right panel of Fig.~\ref{fig:corr-fluc} where fitting using the finite size formula\,\cite{song10}
$$
{\cal F}(r)=\frac{K}{2\pi^2}\ln\left(\frac{L}{\pi}\sin(\pi r/L)\right) + {\rm const.},
$$
leads to much more precise estimate of $K=2.03(2)$, much less sensitive to the fitting process. Given the fact that the numerical computation of ${\cal F}$ is twice less numerically expensive than $G(r)$, this specific example clearly tells us that our method is superior to existing ones. We further demonstrate this fact in the following where we present a quantitative comparison with several previous studies, performed over the last 20 years.
\begin{widetext}
\begin{figure*}[ht!]
\centering
\includegraphics[width=2\columnwidth,clip]{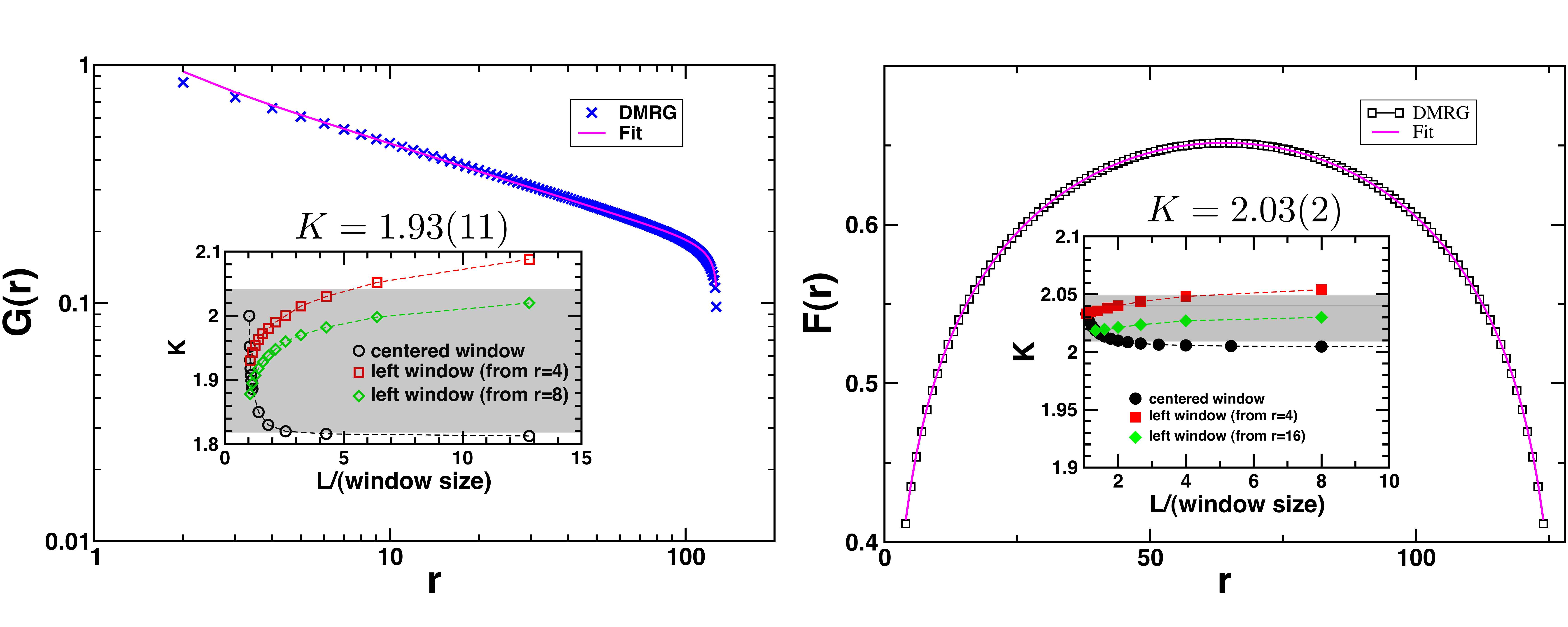}
\caption{DMRG results for the Bose--Hubbard chain at $t/U=0.3$ for $L=128$ (OBC) is shown. Retaining the same number of DMRG states (300), the required CPU time to compute the Gree's function $G(r)$ (Left) was twice larger that for the bipartite fluctuations ${\cal F}(r)$ (Right). While extracting $K$ from the correlations is very sensitive to the fitting window/range (yielding $K=1.93(11)$), it is much less sensitive when extracting $K$ from the fluctuations $\mathcal{F}$ (yielding $K=2.03(2)$).}
\label{fig:corr-fluc}
\end{figure*}

\section{Critical point estimates of the SF-MI transition in the 1D Bose-Hubbard model at unit filling}

\begin{table}[htdp]

\begin{tabular}{|c|c|c|c|l|}\hline
Year&Reference&Technique&Observable&Estimate\\
\hline\hline
1991&Krauth \cite{Krauth91} & (approximate) Bethe Ansatz &&$1/(2\sqrt{3})\simeq 0.2887$\\
\hline
1992&Batrouni {\it et al.} \cite{Batrouni92} & QMC & Superfluid stiffness &$0.2100(100)$\\
\hline
1994&Elesin {\it et al.} \cite{Elesin94} & Exact Diagonalization & Gap &$0.2750(50)$\\
\hline
1996&Kashurnikov {\it et al.} \cite{Kas96a} & QMC & Gap & $0.3000(50)$\\
\hline
1999&Elstner {\it et al.} \cite{Elstner99}& Strong coupling & Gap & $0.2600(100)$\\
\hline
2000& K\"uhner {\it et al.} \cite{Kuhner00} & DMRG & Correlation function & $0.2970(100)$\\
\hline
2008&Zakrzewski {\it et al.} \cite{Zak08} & Time Evolving Block Decimation & Correlation function& $0.2975(5)$\\
\hline
2008 & La\"uchli {\it et al.} \cite{Lauchli08}& DMRG & von Neuman entropy &$0.2980(50)$\\
\hline
2008 & Roux {\it et al.} \cite{Roux08}& DMRG & Gap &$0.3030(90)$\\
\hline
2011& Ejima {\it et al.} \cite{Ejima11} & DMRG & Correlation function &$0.3050(10)$\\
\hline
2011 & Danshita {\it et al.} \cite{Danshita11} & Time Evolving Block Decimation & Excitation spectrum &$0.3190(10)$\\
\hline
{\red{2011}}&{\red This work}& {\red DMRG}&{\red Bipartite Fluctuations}&{\red $0.2989(2)$}\\
\hline
\hline
\end{tabular}
\end{table}
\end{widetext}
\begin{figure}[!ht]
\begin{center}
\includegraphics[width=.8\columnwidth,clip]{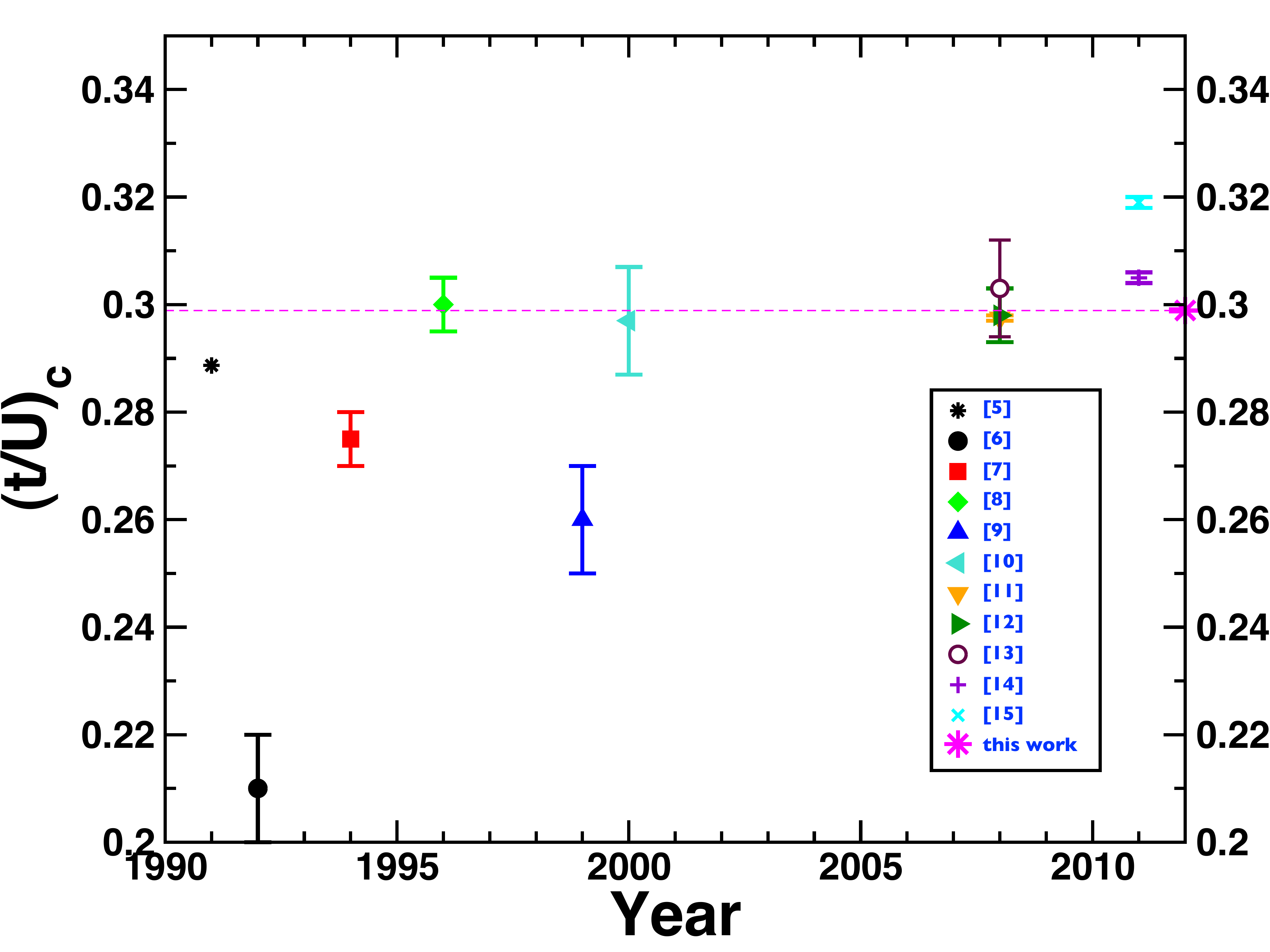}
\caption{Various critical point estimates of the SF-MI transition in the 1D Bose-Hubbard model at unit filling. The plotted values in function of time refers to the above table and to the reference list below.}
\end{center}
\end{figure}


\begin{thebibliography}{10}

\bibitem{vojta}
M. Vojta, Rep. Prog. Phys. {\bf 66},  2069 (2003).

\bibitem{osterloh-02n608-amico-08rmp517}
A. Osterloh, L. Amico, G. Falci, and R. Fazio, Nature {\bf 416},  608  (2002);
L. Amico, R. Fazio, A. Osterloh, and V. Vedral, Rev. Mod. Phys. {\bf 80},  517
  (2008).

\bibitem{Holzhey94-vidal-03prl227902-Calabrese04}
C. Holzhey, F. Larsen, and F. Wilczek, Nuclear Physics B {\bf 424},  443
  (1994); G. Vidal, J.~I. Latorre, E. Rico, and A. Kitaev, Phys. Rev. Lett. {\bf 90},
  227902  (2003); P. Calabrese and J. Cardy,  J. Stat. Mech. P06002  (2004).


\bibitem{hastings07jsmp08024-eisert-10rmp277}
M.~B. Hastings, J. Stat. Mech.  P08024  (2007);
J. Eisert, M. Cramer, and M.~B. Plenio, Rev. Mod. Phys. {\bf 82},  277  (2010).

\bibitem{dalmonte}
M. Dalmonte \ea, Phys. Rev. B {\bf 84}, 085110 (2011).

\bibitem{alet-10prb094452}
F. Alet, I.~P. McCulloch, S. Capponi, and M. Mambrini, Phys. Rev. B {\bf 82},
  094452  (2010).

\bibitem{kallin-09prl117203-hastings-10prl157201}
A.~B. Kallin, I. Gonzales, M.~B. Hastings, and R.~G. Melko, Phys. Rev. Lett.
  {\bf 103},  117203  (2009);
M.~B. Hastings, I. Gonzalez, A.~B. Kallin, and R.~G. Melko, {\it{ibid}}
  {\bf 104},  157201  (2010).

\bibitem{song-11prb224410}
H.~F. Song, N. Laflorencie, S. Rachel, and K. {Le Hur}, Phys. Rev. B {\bf 83},
  224410  (2011).

\bibitem{Alet07} F. Alet, S. Capponi, N. Laflorencie and M. Mambrini, 
Phys. Rev. Lett. {\bf 99}, 117204 (2007).

\bibitem{Gioev06}
D. Gioev and I. Klich, Phys. Rev. Lett. {\bf 96},  100503  (2006).

\bibitem{song-10prb012405}
H.~F. Song, S. Rachel, and K. {Le~Hur}, Phys. Rev. B {\bf 82},  012405  (2010).

\bibitem{song-11}
H.~F. Song, S. Rachel, C. Flindt, I. Klich, N. Laflorencie, and K. Le~Hur, Phys. Rev. B {\bf 85}, 035409 (2012).

\bibitem{suppl}
See Supplemental Material for comparison of the presented approach to other methods.

\bibitem{okamoto-92pla433-eggert96prbR9612}
K. Okamoto and K. Nomura, Phys. Lett. A {\bf 169},  433  (1992); S. Eggert, Phys. Rev. B {\bf 54},  R9612  (1996).

\bibitem{haldane82prb4925}
F.~D.~M. Haldane, Phys. Rev. B {\bf 25},  4925  (1982).

\bibitem{Laflo01} N. Laflorencie, S. Capponi, and E. S. S\o rensen, Eur. Phys. J. B {\bf 24}, 77 (2001).

\bibitem{thomale-10prl116805}
R. Thomale, D.~P. Arovas, and B.~A. Bernevig, Phys. Rev. Lett. {\bf 105},
  116805  (2010).

\bibitem{roncaglia08}
M. Roncaglia \ea, Phys. Rev. B {\bf 77}, 155413 (2008).

\bibitem{Thesberg11} M. Thesberg and E. S. S\o rensen, arXiv:1110.0353

\bibitem{Giamarchi04}
T. Giamarchi, {\em Quantum Physics in One Dimension} (Oxford University Press,
  Oxford, 2004).
  
\bibitem{kuhner-98prb14741}
T.~D. K\"uhner and H. Monien, Phys. Rev. B {\bf 58},  R14741  (1998).

\bibitem{kuhner-00prb12474}
T.~D. K\"uhner, S.~R. White, and H. Monien, Phys. Rev. B {\bf 61},  12474
  (2000).

\bibitem{bhm1d}
V.~A. Kashurnikov, A.~V. Krasavin and B.~V. Svistunov, JETP Lett. {\bf 64},
99 (1996); S. Ejima, H. Fehske and F. Gebhard, EPL {\bf 93}, 30002 (2011);
I. Danshita and A. Polkovnikov, Phys. Rev. A {\bf 84}, 063637 (2011).
  

\bibitem{giamarchi-92prb9325}
T. Giamarchi and A.~J. Millis, Phys. Rev. B {\bf 46},  9325  (1992).

\bibitem{Matsumoto02}
M. Matsumoto, C. Yasuda, S. Todo, and H. Takayama, Phys. Rev. B {\bf 65},
  014407  (2002).

\bibitem{Wenzel08}
S. Wenzel, L. Bogacz, and W. Janke, 
Phys. Rev. Lett. {\bf 101},  127202 (2008).

\bibitem{comment}
To overcome finite-$T$ effects and get ground-state estimates, QMC simulations were performed at $\beta/L=10 J$.

\bibitem{Wang06}
L. Wang, K. Beach, and A. Sandvik, Phys. Rev. B {\bf 73},  014431  (2006).

\bibitem{Bernardet02} K. Bernardet, G. G. Batrouni, J.-L. Meunier, G. Schmid, M. Troyer, and A. Dorneich, Phys. Rev. B {\bf 65}, 104519 (2002).

\bibitem{exp-coldatom}
W.~S. Bakr \ea, Science {\bf 329}, 547 (2010); J.~F. Sherson \ea, Nature {\bf 467}, 68 (2010).

\bibitem{Sandvik07} A. W. Sandvik, Phys. Rev. Lett. {\bf 98}, 227202 (2007).

\end{thebibliography}

\begin{thebibliography}{0}

\bibitem{long}
H.~F. Song, S. Rachel, C. Flindt, I. Klich, N. Laflorencie, and K. Le~Hur, \href{http://prb.aps.org/abstract/PRB/v85/i3/e035409}{Phys. Rev. B {\bf 85}, 035409 (2012)}.

\bibitem{wang06}
Ling Wang, K.~S.~D. Beach, and A.~W. Sandvik, \href{http://prb.aps.org/abstract/PRB/v73/i1/e014431}{Phys. Rev. B 73, 014431 (2006)}.

\bibitem{cazalilla04} 
A. Cazalilla, \href{http://iopscience.iop.org/0953-4075/37/7/051}{J. Phys. B: AMOP {\bf 37}, S1-S47 (2004)}.

\bibitem{song10}
H.~F. Song, S. Rachel, and K. Le~Hur, \href{http://prb.aps.org/abstract/PRB/v82/i1/e012405}{Phys. Rev. B {\bf 82}, 012405 (2010)}.

\bibitem{Krauth91} W. Krauth, \href{http://prb.aps.org/abstract/PRB/v44/i17/p9772_1}{Phys. Rev. B {\bf 44}, 9772 (1991)}.
%
\bibitem{Batrouni92} G. G. Batrouni and R. T. Scalettar, \href{http://prb.aps.org/abstract/PRB/v46/i14/p9051_1}{Phys. Rev. B {\bf 46}, 9051 (1992)}.
%
\bibitem{Elesin94} V. F. Elesin, V. A. Kashurnikov, and L. A. Openov, \href{http://www.jetpletters.ac.ru/ps/1323/article_20004.shtml}{Pis'ma Zh. Eksp. Teor. Fiz. {\bf 60}, 174 (1994) [JETP
Lett. {\bf 60}, 177 (1994)]}.

\bibitem{Kas96a} V. A. Kashurnikov, A. V. Krasavin and B. V. Svistunov, 
\href{http://www.springerlink.com/content/e073436828771uhx/}{Pis'ma Zh. Eksp. Teor. Fiz. {\bf 64}, 92 (1996)
[JETP Lett. {\bf 64}, 99 (1996)]}.
%
\bibitem{Elstner99} N. Elstner and H. Monien, \href{http://prb.aps.org/abstract/PRB/v59/i19/p12184_1}{Phys. Rev. B {\bf 59}, 12184 (1999)}.
%
\bibitem{Kuhner00}
T.~D. K\"uhner, S.~R. White, and H. Monien, \href{http://dx.doi.org/10.1103/PhysRevB.61.12474}{Phys. Rev. B {\bf 61},  12474
  (2000)}.
  %
\bibitem{Zak08} J. Zakrzewski and D. Delande, \href{http://dx.doi.org/10.1063/1.3046265}{AIP Conf. Proc. {\bf 1076}, 292 (2008)}.
%
\bibitem{Lauchli08} A. M. L\"auchli and C. Kollath, \href{http://iopscience.iop.org/1742-5468/2008/05/P05018/}{J. Stat. Mech. (2008) P05018}.
%
\bibitem{Roux08} G. Roux, T. Barthel, I. P. McCulloch, C. Kollath, U. Schollw\"ock, and T. Giamarchi, \href{http://pra.aps.org/abstract/PRA/v78/i2/e023628}{Phys. Rev. A {\bf 78}, 023628 (2008)}.
%
\bibitem{Ejima11} S. Ejima, H. Fehske and F. Gebhard, \href{http://iopscience.iop.org/0295-5075/93/3/30002/}{EPL {\bf 93}, 30002 (2011)}.
%
\bibitem{Danshita11} I. Danshita and A. Polkovnikov,  \href{http://pra.aps.org/abstract/PRA/v84/i6/e063637}{Phys. Rev. A {\bf 84}, 063637 (2011)}.

\end{thebibliography}
\end{document}